# A Visual Analysis Approach to Update Systematic Reviews

Katia Romero Felizardo[1], Elisa Yumi Nakagawa[2], Stephen G. MacDonell[3], and José Carlos Maldonado[4]

[1,2,4]Dept. Computer Systems University of São Paulo São Carlos, Brazil, email: {katiarf, elisa, jcmaldon}@icmc.usp.br,
[3]School of Business University of Otago Dunedin, New Zealand, email: stephen.macdonell@otago.ac.nz

**Abstract**

***Context:*** *In order to preserve the value of Systematic Reviews (SRs), they should be frequently updated considering new evidence that has been produced since the completion of the previous version of the reviews. However, the update of an SR is a time consuming, manual task. Thus, many SRs have not been updated as they should be and, therefore, they are currently outdated.* ***Objective:*** *The main contribution of this paper is to support the update of SRs.* ***Method:*** *We propose USR-VTM, an approach based on Visual Text Mining (VTM) techniques, to support selection of new evidence in the form of primary studies. We then present a tool, named Revis, which supports our approach. Finally, we evaluate our approach through a comparison of outcomes achieved using USR-VTM versus the traditional (manual) approach.* ***Results:*** *Our results show that USR-VTM increases the number of studies correctly included compared to the traditional approach.* ***Conclusions:*** *USR-VTM effectively supports the update of SRs.*

## 1. INTRODUCTION

Evidence Based Software Engineering (EBSE) was first introduced in 2004 as a means of advancing and improving the discipline of Software Engineering (SE) [22]. In this con- text, the Systematic Review (SR) (a.k.a. Systematic Literature Review (SLR)) has provided a methodical, structured process to support the conduction of literature reviews [26] and has gained substantial importance [37]. The reasons why more and more SRs have been conducted every year can be attributed (in part, at least) to the advantages of SRs, including reduced likelihood of bias in results and the potential ability to combine data and aggregate evidence from various quantitative studies, through techniques such as meta-analysis. SRs should provide the available and most up-to-date evidence on a specific topic of interest or research area. However, the conduct of SRs is rather labour intensive in comparison to an "ordinary" narrative literature review, posing several challenges, such as the search for relevant studies [8, 11, 25] and selection of primary studies [12, 13, 23, 32].

If a research area is continually evolving (as is common in computing), SRs that are not maintained (i.e., updated) can become out of date or misleading. Incorporation of new research or evidence into existing SRs is therefore paramount in order to sustain their relevance. In other words, SRs should be frequently updated with the purpose of identifying new evidence that has emerged after the completion of the review. It is worth highlighting that, in the Medicine area, SRs are in general updated at least every two years, determining whether or not there are new studies available for inclusion in a previous review [19]. In the SE area, the main reasons to update SRs is that SE professionals and researchers may rely on the results of SRs to build a body of knowledge about when, how, and in what process, techniques, methods, tools and others are more appropriate to be used. Besides that, SRs in the SE area have also con- tributed to identify new, important research topics that have not been treated yet. Therefore, we argue that the update of SRs is also a quite important issue in SE. However, even when the same authors update their reviews, search and selection of new evidence can take considerable time, especially when many new studies are returned during the search task. Consequently, this leads to difficulties in reading and evaluating the state of the art of a current topic of interest. Thus, it may be beneficial to have approaches, including techniques and tools, which support the update of SRs.

One such approach that we propose here utilises Text Mining. Text Mining is a well-established research area commonly used to extract patterns and non-trivial knowledge from unstructured documents or textual documents writ- ten in a natural language [35]. More specifically, Visual Text Mining (VTM) is an approach that combines mining algorithms and information visualisation techniques to sup- port visualisation and interactive data exploration [36]. In general, humans possess strong visual processing abilities; therefore, visualisation techniques leverage such abilities to support knowledge discovery [21].

Given the above, the main goal of this work is to present an approach, named USR-VTM, to support the update of SRs.

USR-VTM is based on VTM techniques, and so exploits the visual processing abilities of humans. We also present an automated tool, named Revis, which provides support to our approach. In order to evaluate our approach, we have con- ducted an experiment comparing USR-VTM with the traditional, manual approach. The remainder of this paper is organized as follows. Section 2 presents the background and an overview of related work. Section 3 details the proposed approach. Section 4 presents the evaluation of our approach. Section 5 presents a brief discussion of our findings. Finally, conclusions are presented in Section 6.

## 2. BACKGROUND AND RELATED RESEARCH

This section presents brief background information on SRs in the context of SE, as well as on how VTM has been already used in the context of SRs.

### 2.1 SRs in Software Engineering

In general, bibliographical or informal literature reviews do not use a systematic approach. Hence, one cannot rule out that the selection of studies and the conclusions drawn could be biased. This can provide readers with a distorted view about the state of knowledge regarding the area at the focus of the review. Instead, an SR could be used as a means of identifying, evaluating, and interpreting all available re- search relevant to a particular research question, topic area, or phenomenon of interest [26]. The SR uses a systematic process aimed at providing reliable answers to particular re- search questions, using a predefined search, extraction and aggregation strategy.

Several studies have revealed that the SR has become an important research methodology in SE by systematically aggregating evidence on research topics [6, 24, 27, 37]. This growing interest has motivated the establishment of guide- lines, templates, and processes for SR [3, 22]. Kitchenham [22] proposed the first comprehensive guidelines to perform SRs in SE. In addition, Biolchini et al. [3] developed a template that facilitates SR planning and execution. Kitchen- ham [22] also proposed a three-phase process for applying SR: planning, conducting, and reporting. In short, during the planning phase, the goal of the review is identified and a review protocol is developed. In the second phase, identification and selection of primary studies based on the inclusion and exclusion criteria are performed, as well as the review of the selection, quality assessment of primary studies, and data extraction. Finally, the third phase includes data synthesis and reporting of results, to be disseminated to interested parties including researchers and practitioners.

In terms of ensuring the ongoing relevance, two studies have proposed processes for updating SRs [9, 17]. In essence, these processes are composed of three phases: planning, re- view execution, and analysis of results. During the planning phase, the review protocol is revisited in order to identify required changes to make it suitable for updating the SR. For example, the exclusion criteria requires changes to re- strict the search only for studies published after the previous review. The second phase includes discarding of primary studies that overlap with the set of studies retrieved in the previous review. This phase also includes similar activities of the SR process and, in particular, during data extraction, information extracted from studies selected in the new review must be merged with information of the previous review. Finally, the third phase also includes the same activities of the SR process. In spite of these initiatives, in general, when an update is performed, the SE community has still used an ad-hoc approach. Some updated SRs are presented in [2, 6, 10, 27, 34]. In addition, they have been manually updated, without automated support.

### 2.2 Visualization Techniques in the SR Process

The process used to extract high-level knowledge from low-level data is known as Knowledge Discovery in Databases (KDD) [21]. Data Mining (DM) is a part of the KDD process responsible for extracting patterns or models from data. Visual data mining (VDM) is a combination of visualisation and traditional DM techniques used to explore large datasets [7, 21]. A specific application of VDM, which is of interest in our work, is the amalgamation of text processing algorithms with interactive visualisations in order to support users to make sense of text collections. By extension, Visual Text Mining (VTM) refers to VDM applied in text or to a collection of documents [21, 28]. According to research conducted by Paulovich and Minghim [30], the use of VTM can speed up the process of interpreting and extracting useful information from document collections.

Three studies (involving the authors of this work) have investigated the use of VTM within the context of EBSE [14, 15, 29]. Malheiros et al. [29] applied content-based VTM techniques to support the selection of primary studies. Similarly, the approach presented by Felizardo et al. [15] also used VTM techniques in the SR process; however, this approach contains additional visualisation techniques based on meta-data analysis. Malheiros et al. and Felizardo et al. compared the performance of reviewers in carrying out the selection of primary studies by reading abstracts or by using VTM techniques. Both works concluded that VTM provides a more precise selection of relevant studies, speeding up the selection task. Felizardo et al. [14] then extended their previous work [15] in order to support the selection of primary studies, evaluating the decisions of including or excluding primary studies and, mainly, helping reviewers to ensure as far as possible that important studies are not removed. They concluded that VTM can give solid clues about which particular studies should be checked, reducing the volume of documents that need to be re-evaluated and the time spent in the whole process. Practical use of VTM in the context of SRs can be found elsewhere [4]. VTM therefore appears to be beneficial in supporting the SR process. In spite of the relevance of VTM to SRs, it is important to note that its use to update SRs has not been investigated yet.

## 3. USR-VTM

USR-VTM is an approach that supports the update of SRs. As illustrated in Figure 1, our approach comprises three phases: Planning Update, Review Execution, and Analysis of Results. The first and last phases are similar to those presented in previous works [9, 17] (also described in Section 2) and they are not affected by using VTM. Specially, our approach focuses on the second phase (Phase 2: Review Execution). Regarding this phase, it is worth highlighting that one of the most challenging effort and time-consuming tasks in an SR is the selection of new primary studies [32] (i.e., Stage 2.1 in Figure 1). In order to facilitate this task, USR-VTM proposes to use VTM techniques. In the next subsections, we focus on describing how such techniques are used in this phase. Following, we present Revis, a tool that supports our approach.

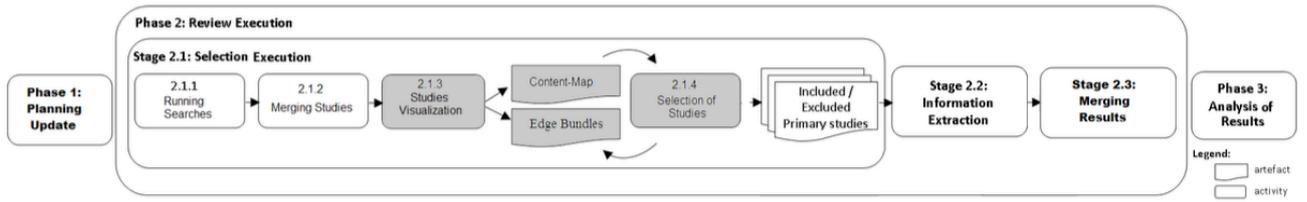

Figure 1: An Overview of USR-VTM.

**3.1 Studies Visualization**

After running the search for primary studies and merging the two sets (i.e., discarding the overlap among results considering the set of studies retrieved in the previous review (Stages 2.1.1 and 2.1.2 of the USR-VTM process – see Figure 1), two visual representations of the aggregated set of primary studies are generated: a content-map and edge bundles (Stage 2.1.3 of the USR-VTM process – see Figure 1). A content-map is a visual representation that aims at supporting the analysis of a set of documents, through similarity of their content. To create this map, three steps are required:

- **Text preprocessing:** This step is responsible for structuring and cleaning data. It receives the set of primary studies as input, including: (i) studies included in the previous review; (ii) studies excluded in the previous review; and (iii) studies to be evaluated in the current review (i.e., "the new evidence"). To represent these studies, we construct a document containing the title, abstract, and keywords of each study. It is worth high- lighting that considering only the title, abstract, and keywords is sufficient because: (i) Dieste and Padua [11] conducted an experiment to analyze if the strategy of searching titles and abstracts is appropriate for use in SRs. Their results confirmed that searching titles and abstracts rather than the full text is a better strategy; and (ii) there are many challenges to manipulate full-text articles, for example, the recognition and clean-up of embedded tags, non-ASCII characters, tables, and figures, and even the need to convert from PDF into textual format [5]. This step converts each document into a vector representation, known as a bag of words [33], based on all the terms extracted from the title, abstract and keywords of primary studies. The initially high number of terms is then reduced. Common non-discriminating terms, known as stopwords, such as prepositions and conjunctions, are removed. Additionally, remaining terms are reduced to their radical using Porter's stemming algorithm [31]. A matrix of documents x terms is built, that compounds the collection's vector representation, where columns rep- resent relevant terms and rows are the frequencies of each term weighted according to the term frequency- inverse document frequency measurement [18]. This measurement makes the importance (weight) of a term directly proportional to its frequency in each document, and inversely proportional to its frequency in the set of documents;

- **Similarity calculation:** After the vector representation is built, the similarity between documents is calculated as the distances between the vectors representing them. For this step, a vector-based distance function can be used. Although any vector-based distance function can be applied, USR-VTM suggests to use a function based on the cosine, which has been shown to be more suitable for SRs [30, 36]. Here we use the common formulation $d(x_i, x_j) = 1 - \cos(x_i, x_j)$, where $x_i$ and $x_j$ are the vectors representing the $i^{th}$ and $j^{th}$ documents, respectively. The cosine similarity between two vectors is a measure that calculates the cosine of the angle between them;

- **Projection:** The final step to obtain the content-map is the projection, which maps each document as a point in a 2D or 3D space. This is achieved through point placement or multidimensional projection techniques [30]. Considering that the vector representation $X = \{x_1, x_2, ..., x_n\}$ is embedded into an $R^m$ space, a multidimensional projection technique (or simply projection technique) can be viewed as a function that maps each m-dimensional instance into a $p$-dimensional instance, with $p = \{2, 3\}$, preserving as much as possible the reduced visual space similarity relationships defined in $\mathcal{R}^m$. In other words, it is a function $f : \mathcal{R}^m \rightarrow \mathcal{R}^p$, which seeks to make $|d(x_i, x_j) - \hat{d}(f(x_i), f(x_j))| \approx 0, \forall x_i, x_j \in X$, where $\hat{d}$ is the distance function in the $p$-dimensional visual space.

The outcome of a projection is a two-dimensional or three-dimensional visual representation (i.e., the content- map), where each m-dimensional instance – in our case, a document (primary study) – is mapped on the screen as a graphic element, normally a circle (as presented in Figure 2(a)). Primary studies with similar content are mapped close to each other and dissimilar ones are positioned far apart. Using the content-map, reviewers can find groups of similar primary studies and can also extract information that are sometimes difficult to attain by only reading the primary studies (particularly if they are read over an extended time period).

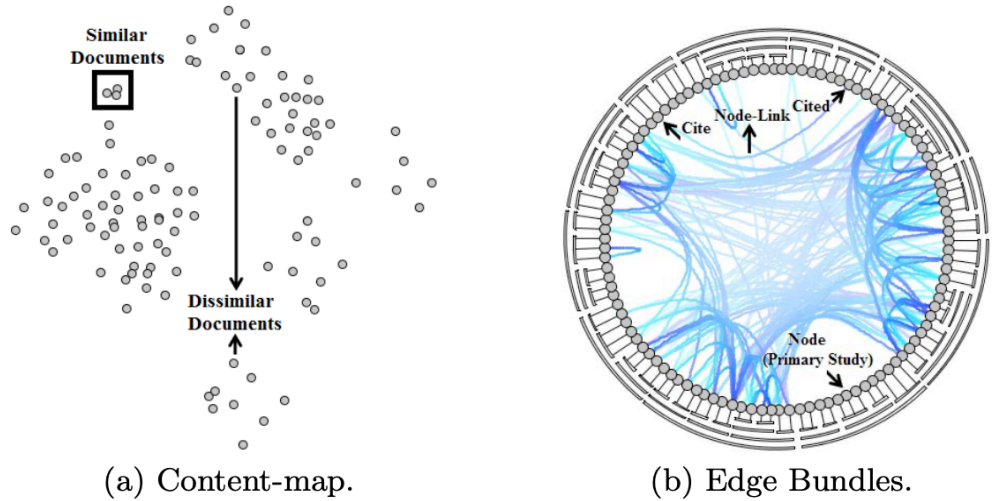

(a) Content-map.  (b) Edge Bundles.

Figure 2: Examples of visualisations to support the SR update process.

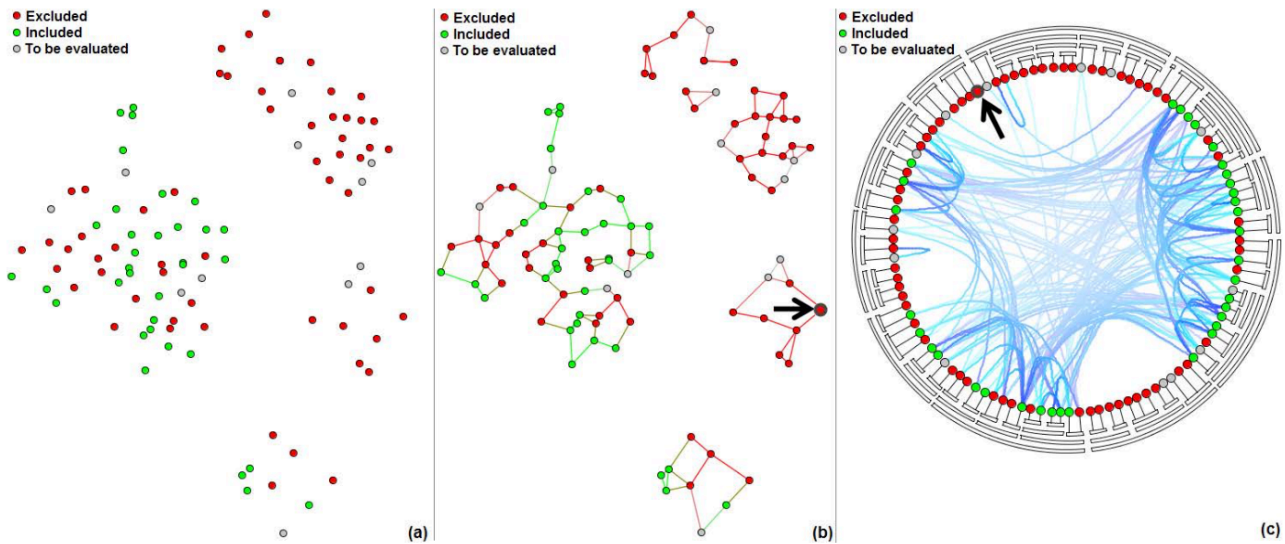

Figure 3: (a) Original content-map; (b) Content-map after application of KNN Edges Connection technique; (c) Edge bundles.

Another VTM technique that supports our approach is edge bundle[1], which is a hierarchical tree visualisation technique that shows both nodes and node-links (relationships between nodes) [20]. In the case of an SR, the nodes (small circles, as shown in Figure 2(b)) are the primary studies (i.e., studies included in the previous review, studies excluded in the previous review, and studies to be evaluated in the current review) and the node-links (blue lines, also shown in Figure 2(b)) are the citations between them. In order to create the hierarchical tree, which is represented by outer circles in the bundles, we have used the HiPP (Hierarchical Point Placement) strategy [30]. Node-links are colored to represent the direction of the citation: the citing primary study is at the light blue end of the link and the cited primary study at the dark blue end. In summary, the edge bundle is used to visualize the number of times that a primary study has been cited by other primary studies.

In next section, we present how the content-map and edge bundles are used to support the selection of new primary studies.

### 3.2 Selection of Studies

In order to select new primary studies using the two visual representations, two steps are performed (Stage 2.1.4 of the USR-VTM process – see Figure 1):

• **Step 1:** The content-map (presented in Figure 3(a)), contains studies included (green points) and excluded (red points) in the previous review, as well as studies to be evaluated (grey points) in the current review. Primary studies are connected with their neighbors, by applying KNN (K-Nearest Neighbor) Edges Connection technique [1] (see Figure 3(b)). This technique connects nodes with their nearest neighbors, computing the proximity on the projection itself.

It is worth noting that studies present in the content-map are also present in the edge-bundles; i.e., each

---

[1] In general, visualisation techniques use colors in order to add extra information on a visual representation. Thus, we suggest reading of a color version of this paper.

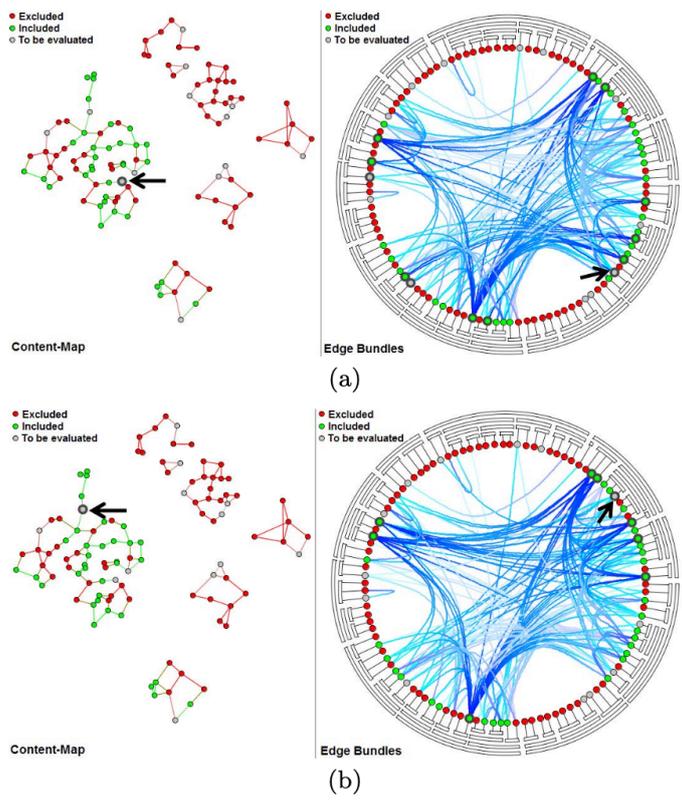
(a)

(b)

Figure 4: Examples of application of strategy 1.

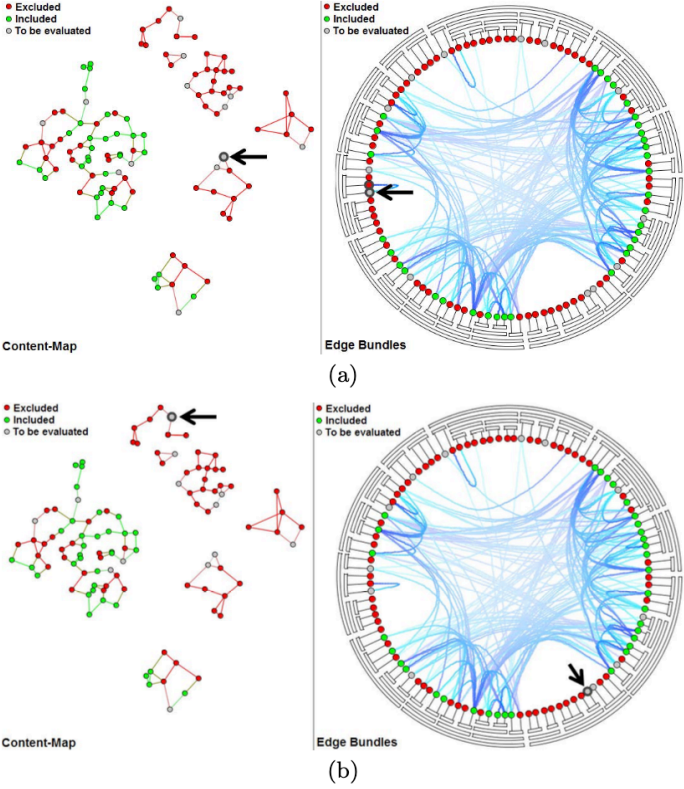
(a)

(b)

Figure 5: Examples of application of strategy 2.

study in the content-map has a corresponding study in the edge bundles, as illustrated by an arrow in Figures 3(b) and 3(c).

• **Step 2:** Two strategies to include and exclude new primary studies using content-map and edge-bundles are then applied:

– **Inclusion Strategy**: To include a primary study, it must be a neighbor of <u>at least</u> one previously included study (observed in the content-map) **AND** it must not cite previously excluded paper(s) (observed in the edge-bundles). Neighbors are nodes connected through edges between them.

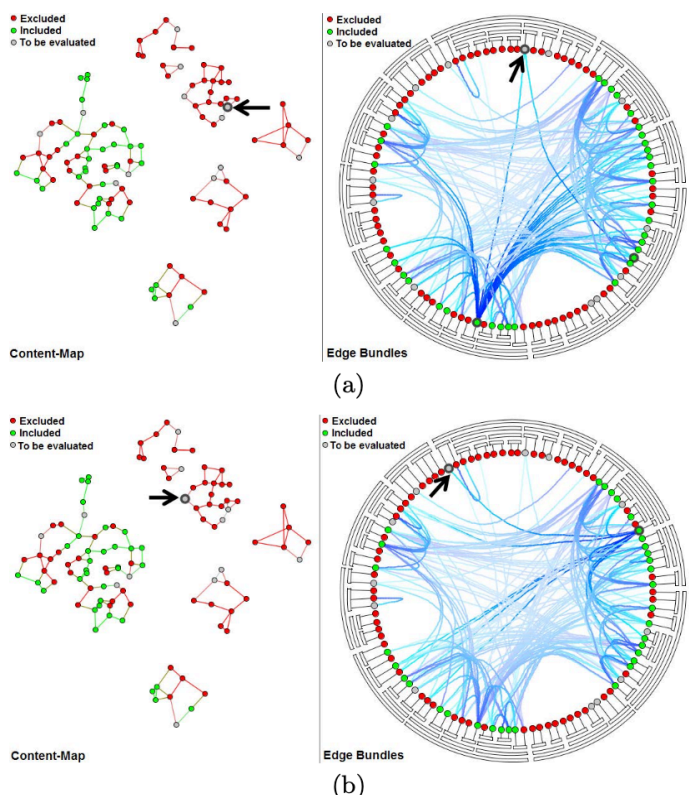

Figure 6: Undefined situations: Include or exclude the new evidence?

Figure 4(a) and Figure 4(b) show two examples of studies that could be included. In both cases, the "New Evidence – NE" (grey points highlighted using arrows) are neighbors of at least one previously included study (green points in the content-map) **AND** they do not cite previously excluded papers (red points in the edge bundles). In or- der to facilitate the identification of cited studies, these studies are presented with a grey circle around them. Thus, both studies should be included in the updated version of the systematic review.

– **Exclusion Strategy:** To exclude a primary study, it must be neighbor of only previously excluded studies (or studies in evaluation) **AND** it must not cite previously included papers.

Figure 5(a) and 5(b) show two examples of studies that could be excluded. In both cases, the "New Evidence – NE" (grey points highlighted using arrows) are neighbors of previously excluded studies (red points in the content-map) **AND** they are not linked to previously included papers (green points in the edge bundles). Thus, they should be excluded.

These strategies were established based on our practical experience acquired over several years, observing several SRs that have been conducted and applying VTM techniques on them. In particular, we have observed the outcomes of SRs when updating them using a traditional approach and using VTM. We emphasise that this experience was obtained in the context of our research group as well as through international collaborations.

Undefined or unclear situations should be given particular attention. For example, a study may be a neighbor of previously excluded studies; however, it cites previously included study(ies), as illustrated in Figure 6. The "New Evidence – NE" (grey points indicated by arrows in both views) are linked to previously excluded study(ies) (red point(s) in the content-map) and cite previously included paper(s) (green point(s) in the edge bundles). In a situation such as this, the new primary study should be analyzed by the reviewer.

### 3.3 Tool Support

The use of VTM techniques requires support of a software tool to be adequately viable. We present our tool, named Revis[2], to automate Stages 2.1.3 (Studies Visualization) and 2.1.4 (Selection of Studies) of our approach. Revis is an open source visualisation and interaction tool that offers a framework of different projection techniques and methods to create content-maps (in particular, based on content similarity) and visualisations based on citation and co-citation relationships amongst documents. Revis takes as input a set of primary studies (included and excluded ones) selected during the previous SR and new primary studies (i.e., "new evidences") found during the update of the SR. These studies are organized according to the bibtex format, which includes title, abstract, keywords, references, and status (i.e., included, excluded, and to be evaluated). Revis then executes the activities performed during Stage 2.1.3 and presents the content-map and edge bundles for the set of primary studies. The studies are automatically colored in these views, according to the previously defined status: studies included in the previous SR are colored in green, studies excluded from the previous SR are colored in red, and the studies to be evaluated are colored in grey.

---

[2] Available in http://ccsl.icmc.usp.br/redmine/projects/revis/files/.

The main functionalities offered by Revis and relevant to our approach are: (i) it creates the views (content-map and edge bundles); (ii) it allows changing of visual attributes (color) of the points (primary studies) in the content-map to represent their status; (iii) it makes it possible for users to explore neighborhood relationships in the content-map, with neighboring connections (called KNN connections) shown as edges among the primary studies; (iv) it supports coordination between the content-map and edge bundles; and (v) it displays the content of a primary study (i.e., title, abstract, and keywords) in a separate window when the user double clicks a specific point.

## 4. EXPERIMENT

In this section, we present an experiment that was performed to evaluate the viability of our USR-VTM approach. Through this experiment, we argue that VTM techniques can facilitate the selection of "new evidence" during the SR update process. Hence, our research questions (RQ) are: (i) **RQ1:** Do VTM techniques (content-map and edge bundles) improve the performance (time taken) of the selection of primary studies during the SR update process?; (ii) **RQ2:** Do VTM techniques improve the accuracy (number of primary studies correctly included or excluded) of the selection of primary studies during the SR update process?

### 4.1 Training and Execution

The subjects involved in this study were 12 graduate students (10 PhD and 2 Master's students) of an SE course at University of São Paulo, Brazil. The experiment was organized in two sessions: training and execution. For training purposes, a set of data (SR Dataset 1, containing 57 included studies, 6 excluded studies, and 15 studies to be analysed) and a set of inclusion and exclusion criteria were used. For the execution session, a different set of data (SR Dataset 2) and a set of inclusion and exclusion criteria were used.

Dataset 2 originated from an SR on aspect-oriented software testing, which was first performed in July, 2007 [16]. In this SR, 97 primary studies were found (where 63 were included and 34 were excluded). The authors updated their SR in September, 2008 [17] and 13 new primary studies were found (where 6 were included and 7 were excluded). These 6 primary studies marked as included by Ferrari et al. [17] (re- viewers who conducted the SR) were taken as oracles, since these reviewers are specialists in both aspect-orientation and software testing. It is worth noting that we chose these relatively small datasets (Dataset 1 and Dataset 2) on the assumption that an experiment with a much greater number of primary studies could negatively affect the motivation and performance of the subjects.

The subjects were split into two groups, each one containing 5 PhD students and 1 Master student, both with prior experience in conducting SRs. During the training session, all subjects received an overview of the experiment and an explanation on their task. Moreover, since specialists in a research topic are normally required in order to conduct an SR about that topic, we also provided extensive training about aspect-orientation and software testing, i.e., the topics of the SR to be used during the execution session, for subjects of both groups. Through this, we aimed to minimize wrong decisions (to include or exclude primary studies) due to a lack of knowledge about the topics involved in the SR. Only subjects from Group 2 were trained on how to use the Revis tool and the VTM techniques. During this training that took approximately 60 minutes, doubts about the tool and VTM techniques were also clarified. The task associated with each group were: **Group 1**– conduct the selection activity using the traditional approach, i.e., reading of abstracts of the 13 new evidence papers (primary studies) from Dataset 2 and mark them as included or excluded in accordance to the inclusion and exclusion criteria; and **Group 2**– conduct the selection activity with VTM support, i.e., analysing the content-map and edge bundles views containing the 63 studies included, 34 excluded, and 13 to be evaluated from Dataset 2 and marking them as included or excluded. It is important to note that the subjects of Group 2 were not permitted to have access the title, abstract, and keywords or any information of the primary studies.

In the execution session, no time limit was imposed on the sub jects to execute the selection of "new evidence", and communication with each other was not allowed. In order to analyse the results of our experiment, we measured performance and accuracy, where *performance* and accuracy were calculated as follow: *Performance = T* , where: T = Time taken for selection of primary studies. *Accuracy = M*, where: M = Number of primary studies correctly selected , i.e., number of selected studies that matched with the oracle.

### 4.2 Results

Table 1 summarizes the results of our experiment. Note that no statistical significance tests were performed due to the very small sample sizes employed. To answer our first research question (RQ1), performance of the subjects was measured (column 3 in Table 1). The time spent by subjects of Group 1 varied between 14 and 53 minutes, while time spent by subjects of Group 2 to perform the same activity using our approach varied between 13 and 25 minutes. The average time of Group 1 was 31.6 minutes (median 24.0 minutes) and the average time of Group 2 was 17.8 minutes (median 16.5 minutes). The standard deviation was 17.7 minutes for Group 1 and 5 minutes for Group 2. The highest standard deviation of Group 1 indicates that the times are spread out over a larger range of values. Therefore, performance in using our proposed USR-VTM approach was better than in traditional approach.

Columns 4 and 5 show, respectively, the number of studies correctly classified (correctly included and correctly excluded) by each subject (i.e., the accuracy). For instance, Subject 1 of Group 1 correctly included 5 studies (from 6 studies that were our oracle) and correctly excluded 4 (from 7 of our oracle). Therefore, 9 (i.e., 5+4) studies were correctly classified, corresponding to 69.2% (i.e., 9/13) of studies for this subject. The accuracy of subjects of Group 1 varied between 7 and 12 (of 13) studies correctly classified, while the accuracy of subjects of Group 2 to perform the same activity using our approach varied between 10 and 13 studies. The average values (and median) of studies correctly classified were 9.5 in Group 1 and 12 studies in Group 2. The standard deviations were 1.64 studies for Group 1 and 1.09 studies for Group 2. A small standard deviation means that the experimental values are clustered together tightly, i.e., the range of values is low and homogeneous. Therefore, ac- curacy using our proposed USR-VTM approach was higher than using the traditional approach.

Columns 7 and 8 present the number of studies incorrectly included and incorrectly excluded, respectively, by each subject. For example, Subject 1 of Group 1 included 3 studies that should have been excluded and excluded 1 study that should have been included. This reviewer classified incorrectly 4 (i.e., 3+1) studies. Subjects of Group 1 incorrectly classified from 1 to 6 studies, while subjects of Group 2 incorrectly classified from 0 to 3 studies. The average (and median) value of studies incorrectly classified in Group 1 was 3.5 studies, whereas in Group 2 it was only 1 study. The standard deviations were 1.64 studies for Group 1 and 1.09 studies for Group 2. A small standard deviation mean the values in a statistical data set are close to the mean of the data set, on average. Therefore, the number of studies in- correctly classified using our proposed USR-VTM approach was lower than using the traditional approach.

Our small samples precluded formal statistical testing of the differences in performance and accuracy. The results do pro- vide a degree of evidence that the use USR-VTM approach may be more effective than a manual update, however. This result could be verified through subsequent larger-scale experiments.

## 5. DISCUSSION

Considering that up-to-date SRs are quite important, but effective update approaches have not been widely investigated in the literature, our work proposes the use of VTM techniques in this scenario. In particular, our approach presents good potential to be used during selection of new primary studies, considering information from an SR that has been previously conducted. Results achieved show that performance of subjects using our approach is on average better when compared to those using a traditional approach. This can be justified since VTM techniques facilitate the extraction of high quality information even from a large amount of data [21], accelerating the rate at which analyses of the high volume of primary studies can be undertaken. Moreover, it is important to highlight that USR-VTM improves accuracy (number of new studies correctly included) in the updated version of the SR in comparison to the traditional approach. In addition, the number of studies incorrectly excluded is lower using USR-VTM. Therefore, USR-VTM can be considered to be a relevant contribution to the area of EBSE. The employed visual representations can be used to support the decisions made by reviewers regarding inclusions and exclusions and to ensure that relevant studies have not been eliminated.

Another contribution of this work is to make available an open source visualisation tool, in the case of Revis. It may be used as a supporting tool by anyone who intends to use USR-VTM. Besides that, if it is interesting, similar tools could be implemented based on Revis, considering that its source code is also available. Using tools such as Revis, which provides a set of well-known, experimented, and used visualisation techniques, interpretation of the views will not require additional training or knowledge.

A list of studies and their respective status (i.e., included in a previous SR, excluded from a previous SR, and to be evaluated) are required to update existing SRs. Duplicated studies are identified manually and the list of studies can be organized according to the bibtex format. However, most academic search engines support the to download studies in this format, favoring the application of our approach in practice.

We have also foreseen the use of USR-VTM as a complementary support to the traditional approach to the SR up- date process. In this perspective, USR-VTM could be used after applying the traditional approach for selecting new primary studies. USR-VTM could suggest primary studies that should be re-checked because the traditional approach included them (and USR-VTM suggests to exclude them) or excluded them (and USR-VTM suggests to include them).

One of the potential threats to the internal validity of our experiment is related to our assumption that the work of Ferrari et al. [17], used as our oracle, correctly included or excluded primary studies. However, we believe that decisions of inclusion or exclusion were correctly performed by these authors, since they are specialists in aspect-orientation and software testing, as well as SR. A second threat is the small size of the subject set in our experiment. However, for a first assessment of our approach, we believe our experiment met its goal. We also acknowledge the potential threats that arise in relation to the selected test subjects, being students at one of our institutions, and the overlap in the authors of this work and the SR used in the experiment. Although our experiment contains a rather small number of studies, the Revis and VTM approach suggested by us can be used in real SRs, where a large number of candidate studies are considered – hundreds and even thousands.

For future work, we intend to conduct other experiments using larger sample sizes of participants, SRs with large amounts of data, SRs performed by other researchers, and reviewers with more experience on the topics of the SRs, in order to increase the reliability of our findings. We also intend to conduct experiments by updating previously con-ducted SRs, and comparing the outcomes of our approach and of traditional one. Another intention is to use USR-VTM to also update Systematic Mappings (SM). Since se-lection of new primary studies is quite similar in both SM and SR, we believe our approach is also suitable to it.

## 6. CONCLUSIONS

Since knowledge in any research area continually evolves, SRs that intend to represent such knowledge must be also frequently updated; otherwise, their validity is undermined.

Table 1: Summary of Results.

| Group | Subject ID | Time (min) | Correctly Included | Correctly Excluded | Total Correctly Classified (%) | Incorrectly Included | Incorrectly Excluded | Total Incorrectly Classified |
|---|---|---|---|---|---|---|---|---|
| Group 1 | 1 | 53 | 5 | 4 | 09 (69.2%) | 3 | 1 | 4 |
| | 2 | 55 | 5 | 7 | 12 (92.3) | 0 | 1 | 1 |
| | 3 | 20 | 4 | 6 | 10 (76.9) | 1 | 2 | 3 |
| | 4 | 22 | 3 | 7 | 10 (76.9) | 0 | 3 | 3 |
| | 5 | 14 | 2 | 7 | 09 (69.2) | 0 | 4 | 4 |
| | 6 | 26 | 1 | 6 | 07 (53.8) | 1 | 5 | 6 |
| Group 2 | 1 | 23 | 5 | 7 | 12 (92.3%) | 0 | 1 | 1 |
| | 2 | 13 | 6 | 7 | 13 (100.0) | 0 | 0 | 0 |
| | 3 | 19 | 5 | 7 | 12 (92.3) | 0 | 1 | 1 |
| | 4 | 25 | 5 | 5 | 10 (76.9) | 2 | 1 | 3 |
| | 5 | 13 | 6 | 7 | 13 (100.0) | 0 | 0 | 0 |
| | 6 | 14 | 5 | 7 | 12 (92.3) | 0 | 1 | 1 |

In this scenario, where most SRs have not been updated or, when updated, considerable efforts must be applied, the main contribution of this work is the USR-VTM approach to support the update process of SRs. USR-VTM is a natural continuation of our previous works [15, 14], where we have been applying VTM techniques and Revis for study selection as well its validation. In particular, our approach in this significant extension leverages the use of VTM techniques in the context of SRs update. By exploiting the strong visual processing abilities of humans, these techniques have proved to be an important ally to improve performance and ac- curacy during the update of SRs, facilitating and enhancing interpretation and decision-making in regard to the selection of new primary studies. Considering the initially good perspectives that we have found in using VTM in the context of SR, it is important to continue our investigation, conducting more experiments, as well as applying our approach to update SRs. Efforts in this perspective could contribute to the EBSE area and also to other research areas that have already discovered the relevance of SR.

## Acknowledgment

The authors would like to thank the Brazilian funding agencies CNPq (Process n. 573963/2008-8 and 474720/2011- 0), Capes (Process n. 034/12), and FAPESP (Process n. 2012/02524-4, 08/57870-9 and 2011/23316-8).